\documentstyle[12pt]{article}
\def\theequation{\ksection.\arabic{equation}}

\def\be{\begin{equation}}
\def\ee{\end{equation}}
\def\bea{\begin{eqnarray}}
\def\eea{\end{eqnarray}}
\def\l{\label}
\def\c{\cite}

\def\log10{{\rm log}_{10}}
\def\massp{M_{\rm Pl}}
\def\G{\Gamma}
\def\COBE{{\em COBE}/DMR }

\begin{document}
\begin{titlepage}

\begin{flushright}
{\small
FERMILAB Pub-94/109-A\\
April 1994 \\
Submitted to {\em Physical  Review D}}
\end{flushright}
\vspace{.35in}

\begin{center}
\Large
{\bf Black Hole Relics and Inflation: Limits on Blue Perturbation Spectra}

\vspace{.25in}

\normalsize

\large{B.J. Carr and J.H. Gilbert$^1$}

\normalsize
\vspace{.5cm}

{\em Astronomy Unit, School of Mathematical Sciences, \\
Queen Mary and Westfield, Mile End Road, London E1 4NS. U.K.}

\vspace{.5cm}

\large{James E. Lidsey$^2$}

\normalsize
\vspace{.5cm}
{\em NASA/Fermilab Astrophysics Center, \\
Fermi National Accelerator Laboratory, Batavia, IL 60510-0500. U.S.A.}
\vspace{.5cm}

\end{center}

\baselineskip=24pt
\begin{abstract}
\noindent

Blue primordial power spectra have spectral index $n>1$ and  arise
naturally in the recently proposed hybrid inflationary scenario. An
observational upper limit on {\em n} is derived by normalizing the
spectrum at the quadrupole scale and considering the possible
overproduction of Planck mass relics formed in the final stage of
primordial black hole evaporation. In the inflationary Universe with
the maximum reheating temperature compatible with the observed
quadrupole anisotropy, the upper limit is $n=1.4$, but it is slightly
weaker for lower reheat temperatures. This limit applies over 57
decades of mass and is therefore insensitive to cosmic variance and
any gravitational wave contribution to the quadrupole anisotropy. It
is also independent of the dark matter content of the Universe and
therefore the bias parameter. In some circumstances, there may be an
extended dust-like phase between the end of inflation and reheating.
In this case, primordial black holes form more abundantly and the
upper limit is $n=1.3$.

\vspace{.5cm}

\noindent
PACS number(s): 98.80.Cq,97.60.Lf,98.70.Vc

\vspace{.1in}

\noindent
\small Electronic address: $^1$J.H.Gilbert@qmw.ac.uk;  \\
Electronic address: $^2$jim@fnas09.fnal.gov

\end{abstract}

\end{titlepage}

\setcounter{equation}{0}
\section{Introduction}

\setcounter{equation}{0}

Is is now generally accepted that large-scale structure in the
Universe arose from the growth of small density fluctuations through
gravitational instability.  A determination of the power spectrum of
these primordial fluctuations is therefore of fundamental importance
and the inflationary scenario \c{inflation} provides an attractive,
causal mechanism for producing such a spectrum (see
\c{LL1993} for a recent review). During inflation the Universe is
dominated by the self-interaction potential $V(\phi)$ of a quantum
scalar field $\phi$ and this results in a superluminal expansion of
the scale factor.  Quantum fluctuations in this field are therefore
stretched beyond the Hubble radius $H^{-1}$, where they remain frozen
until reentry during the radiation- or matter-dominated eras.

In general, when scalar density fluctuations reenter the Hubble
radius, their amplitude is given by
\be
\l{scalar}
\delta \approx  \frac{1}{\massp^3} \frac{V^{3/2}}{|V'|} ,
\ee
where the quantities on the right-hand-side are evaluated when the
scale first crossed the Hubble radius during inflation, a prime
denotes differentiation with respect to $\phi$ and $\massp$ is the
Planck mass \c{BST1983}. The form of these fluctuations can be
parameterized by the spectral index $n$ which specifies the dependence
of the power spectrum on the comoving wave-number $k$
(viz. $|\delta_k|^2 \propto k^n$).  This means that the density
perturbations have an rms amplitude of the form $\delta (M) \propto
M^{(1-n)/6}$ when the mass scale $M$ reenters the horizon after
inflation.

In the simplest scenarios these horizon-scale fluctuations are
expected to be almost scale-invariant with $n \approx 1$ \c{BST1983}.
However, observations of large scale structure suggest that
scale-invariant fluctuations may not work, at least for the cold dark
matter (CDM) model \c{LL1993,saun}, and it may be necessary to
consider `tilted' spectra with $n \ne 1$ \c{LC1992}. `Blue' primordial power
spectra have $n>1$ and are currently consistent with   the recent anisotropy
measurements of the Cosmic
Microwave Background (CMB) radiation [6-13]. A best-fit of the
theoretical and observed autocorrelation functions with the \COBE
first-year maps implies $n=1.15^{+0.45}_{-0.65}$
\c{S1992}.  Smoot et al. \c{S1994} and Torres \c{T1994} have
independently performed a topological analysis of this data and deduce
$n=1.7^{+0.3}_{-0.6}$ and $n=1.2^{+0.3}_{-0.3}$, respectively. On
the other hand, Bond \c{B1994} infers $n=1.8^{+0.6}_{-0.8}$ from FIRS data at
$3.8^o$ \c{FIRS} and $n=2.0^{+0.4}_{-0.4}$ from the first-year \COBE data.
The second-year \COBE data has now been analyzed by Wright et al.
\c{W1994} who find $n=1.46^{+0.41}_{-0.44}$, while a maximum-likelihood
analysis of this data by Bennett et al. \c{Be1994} implies that
$n=1.59^{+0.49}_{-0.55}$. The 53 and 90 GHz second-year data has also
been analyzed by G\'orski et al. \c{G1994}, who find a maximum
likelihood value of $n=1.22 $ $(1.02)$ if the quadrupole is included
(excluded). They further obtain a marginal probability distribution
with a mean of $n=1.10 \pm 0.32$ $(0.87 \pm 0.36)$.\footnote{It is now
recognized, however, that the Doppler peak influences the low multipole
harmonics of the temperature anisotropy in such a way that the observed
spectral index is tilted slightly to the
 blue end, i.e. the spectral index as measured by \COBE is not the index of the
primordial spectrum. For standard CDM the correction is of order $15\%$.} It is
interesting
to note that $n>1$ is deduced by Piran et al.  from
large-scale structure considerations \c{P1993}. If the voids detected
in the CfA survey \c{CfA} on scales $50h^{-1}$ Mpc arise from an
underdensity in the matter distribution and form gravitationally in an
$\Omega =1$ Universe, a spectral index of $n\approx 1.25$ is
consistent with the \COBE detection.

The above results implicitly assume that the spectral index is
constant over the scales of interest, but in general $n$ is a function
of scale and is determined by the magnitude of the potential and its
first and second derivatives \c{LL1992}. However, scales relevant for
large-scale structure lie in the range $1$  - $10^4$ Mpc and correspond
to only $9$ e-foldings of inflationary expansion. Since the scalar
field must roll down its potential very slowly during inflation, only
a very narrow region of $V$ is relevant for these scales. It is
therefore consistent to expand the potential as a Taylor series about
a given scale and this is equivalent to assuming that the spectral
index is constant over a sufficiently small range
\c{reconstruct,Turner}.

The full class of general potentials leading to spectra with constant
spectral index has now been derived \c{CL1993,Lidsey}. The potential
leading to $n>1$ is a combination of trigonometric functions
\c{CL1993} and its Taylor expansion to quadratic order is given by
\be
\l{taylorexp}
V(\phi) \approx V_0 \left[ 1+  2 \pi (n-1) \frac{\phi^2}{\massp^2} \right] ,
\ee
where $V_0$ is a constant that can be  normalized to the \COBE   quadrupole via
Eq. (\ref{scalar}). It should be emphasized that {\em any} potential exhibiting
 a Taylor expansion of this form over scales corresponding to large-scale
structure  leads to a
 blue spectrum. Blue spectra also arise in some variants of the hyperextended
scenario \c{extended}.

Such potentials form the basis of the hybrid inflationary scenario
\c{hybrid} and can arise in string physics \c{Cetal}. The \COBE
detection implies that the last stages of inflation must have occurred
at or below the Grand Unification scale and the superstring is an
effective $N=1$ supergravity theory at these scales. It can be shown
that under fairly generic circumstances the one-loop K\"ahler
potential derived from the orbifold compactification of the
superstring is given approximately by Eq. (\ref{taylorexp}) \c{Cetal},
where the spectral index is determined by fundamental string
parameters:
\be
\l{stringspectral}
n=1+\frac{\delta_3^{\rm GS}}{2\pi^2  \langle S +\bar{S}\rangle} .
\ee
The dilaton field $S$ is the real component of the chiral multiplet
and its vacuum expectation value determines the string coupling
constant $\langle {\rm Re}(S) \rangle \approx g^{-2}_{\rm GUT}$ .
Results from LEP imply that $g^2_{\rm GUT} \approx 2\pi /13$ and
consequently $\langle {\rm Re}(S)\rangle \approx 2$ \c{LEP}. The
parameter $\delta_3^{\rm GS}$ is a dimensionless coefficient to the
one-loop corrections arising in the Green-Schwarz mechanism and
calculations suggest that $\delta_3^{\rm GS} \le 4\pi^2$ \c{suggest}.
These values lead to the upper limit $n\le 1.5$, although the
requirement that inflation ends at the correct time leads to a
stronger limit $n\le 1.3$ \c{Cetal}.

We therefore infer that values of $n>1$ are not excluded at this
stage, either observationally or theoretically, and it is therefore
important to obtain upper limits on the spectral index. Blue spectra
introduce more short-scale power and this might be problematic for
dark matter models of galaxy formation.  However, non-linear effects
become more important on larger scales when $n>1$ and these effects
must be adequately accounted for before such spectra can be excluded
by large-scale structure arguments.  In principle, the extra
short-scale power can be significantly reduced by the free streaming
of a hot dark matter component and comparison of the mixed dark matter
model with observations of large-scale structure above $1$ Mpc implies
an upper limit of $n\le 1.35$ \c{LL1994}, with dependence on the
current value of the Hubble constant. The limit on the spectral
distortion of the CMB from the {\em COBE} FIRAS experiment leads to a
slightly weaker limit of $n<1.56$ \c{FIRAS}.

One of the most interesting constraints on inflationary scenarios
which produce blue spectra comes from considering the formation and
subsequent evaporation of primordial black holes (PBHs) \c{CL1993}.
PBHs are never produced in sufficient numbers to be interesting if
$n<1$, but they could be if $n>1$. Such limits are interesting because
they constrain the power spectrum, and therefore the inflationary
potential, over 47 decades of mass, whereas large-scale structure
measurements span only 10 decades.  This implies that the   limits on the
spectral index derived from PBHs
are extremely {\em insensitive} to the precise value of the
quadrupole anisotropy and so the problems associated with cosmic
variance on these ultra-large scales are evaded. They are also
independent of any gravitational wave contribution to the CMB
anisotropy and, when the spectrum is normalized to the {\em COBE}
detection, they are also independent of the transfer function, the
bias parameter and the value of Hubble's constant. Although such  observational
limits are  essentially independent of observations, they are derived from the
assumption that the spectral index is constant over a very large range of
scales. In the Appendi
x it is shown that this is a valid assumption in the hybrid inflationary
scenario.

In this work we shall consider the potentially stronger observational
constraints which arise if evaporating PBHs leave stable Planck mass
relics.  Several people have considered the cosmological consequences
of such relics. MacGibbon considered the possibility that they could
have around the critical density and thus provide the dark mass
required in galactic halos or the cosmological background
\c{M1987}. She argued that, in the standard non-inflationary scenario,
this would happen rather naturally if the PBHs formed from
scale-invariant fluctuations (n=1) with those of $10^{15}$g having the
density required to contribute appreciably to $100$ MeV cosmic
rays. However, the relic density would be much reduced if there were
an inflationary period. Barrow et al.  have studied this problem in
more detail, calculating the constraints on the fraction of the
Universe going into PBHs in order that their relics do not have {\em
more} than the critical density \c{BCL1992}. They considered the
situation in which PBHs form as a result of a first-order phase
transition induced by bubble collisions at the end of an extended
inflationary period. We go beyond these calculations in several
respects. Firstly, we assume that the PBHs form directly from the
inflation-induced density fluctuations rather than from bubble
collisions. Secondly, we allow for the possibility that the equation of
state may go soft in some period between the end of inflation and
reheating, as first suggested by Khlopov et al. \c{KMZ1985}.

The plan of this paper is as follows. In Section 2 we discuss why PBHs
may form after inflation and why stable relics may be left over from
the final stages of their evaporation. In Section 3 we derive limits
on the fraction of the Universe which goes into PBHs of mass $M$ which
leave relics, both for the standard case in which the Universe is
radiation-dominated after inflation and for the case in which there is
an intermediate dust phase before reheating. In both cases we infer
corresponding upper limits on the amplitude $\delta (M)$. In Section 4
we combine these limits with the observed quadrupole anisotropy to
infer upper limits on $n$. If there were no Planck mass relics, the
strongest upper limit would be associated with the photodissociation
of primordial deuterium by photons emitted from $10^{10}{\rm g}$ PBHs
\c{L1980} and this would imply $n\le 1.5$ \c{CL1993}. If there are
Planck mass relics and there is no dust phase, the limit is
strengthened to $n<1.4$ for a reheat temperature of order
$10^{16}$GeV. If there is an extended dust phase after inflation, the
limit on $n$ depends crucially on its duration and could be as strong
as $n < 1.3$.  In Section 5 we draw some general conclusions.

\section{The Formation of PBHs and their Relics}

\setcounter{equation}{0}

\subsection{The Fraction of the Universe going into PBHs}

We are interested in the situation where PBHs form from the density
perturbations induced by quantum vacuum fluctuations. We assume
spherically symmetric, Gaussian fluctuations with an rms amplitude
$\delta (M)$ and a background equation of state $p=\gamma \rho$ with
$0<\gamma <1$. When an overdense region stops expanding, it must have
a size of at least $\sqrt{\gamma}$ times the horizon size in order to
collapse against the pressure and this requires that $\delta (M)$
exceed $\gamma$ \c{C1975}. Thus the probability of a region of mass
$M$ forming a PBH is
\be
\l{PBHform}
\beta_0 (M) \approx \delta (M) \exp \left( - \frac{\gamma^2}{
2\delta^2(M)} \right)
\ee
and this gives the fraction of the Universe expected to go into PBHs
on that scale. The mass of a PBH forming at time $t$ must be at least
$\gamma^{3/2}$ times the horizon mass, so
\be
\l{softmass}
M = \eta {t\over t_{Pl}} \massp ,
\ee
where $\eta \approx \gamma^{3/2}$. The value of $\gamma$ will usually
be $1/3$ in the early Universe, corresponding to a radiation equation
of state. However, one may have $\gamma =0$ if the Universe ever
passes through a dust-like phase and, in this case, both
Eqs. (\ref{PBHform}) and (\ref{softmass}) are inapplicable
\c{polnarev}. During a dust era, the fraction of the Universe going
into PBHs just depends on the probability that regions will be
sufficiently spherically symmetric to collapse within their
Schwarzschild radius and this can be shown to be \c{polnarev}
\be
\l{betaM}
\beta (M) \approx 2\times 10^{-2} \delta (M)^{13/2} .
\ee
If this applies for some period $t_1<t<t_2$, then PBHs are expected to
form in the mass range $M_1 \le M\le M_{\rm max}$, where $M_1$ is
given in terms of $t_1$ by Eq. (\ref{softmass}) and $M_{\rm max}$ is
the mass of a configuration that just detaches itself from the
universal expansion at $t_2$. The latter is calculated as follows. If
a region has a fluctuation $\delta (M)$ at the time $t_H(M)$ when it
reenters the horizon, then it binds at a time $t_B=t_H\delta^{-3/2}$
with a size $R_B=ct_H\delta^{-1}$. The horizon size at $t_B$ is
$ct_H\delta^{-3/2}$, so $R_B$ is $\delta^{1/2}$ times this and the
maximal mass is given implicitly by \c{polnarev}
\be
\l{maximummass}
M_{\rm max}=  [\delta (M_{\rm max}) ]^{3/2} \left(\frac{t_2}{t_{Pl}}\right)
\massp .
\ee
The last two terms just give the horizon mass at $t_2$. In order to
determine $M_{\rm max}$ explicitly, one needs to know the form of $
\delta(M)$.

The possibility of a soft equation of state is particularly relevant
to the present work because, in some circumstances, this may occur
during the reheating phase at the end of inflation \c{KMZ1985} and
this is precisely the period in which most PBHs are {\em expected} to
form. If inflation ends by means of a second-order phase transition,
the scalar field oscillates in the potential minimum from the time
$t_1\equiv H^{-1}$ and then reheating occurs as friction generated by
the coupling of the scalar field to other matter fields turns the
kinetic energy of motion into background radiation.  The reheating is
completed on a time scale determined by the decay width $\Gamma$ of the
scalar field. Providing $\Gamma^{-1} \ll t_1$, the reheating time can
be taken to be $t_1$, so we can assume that the equation of state is
hard $(\gamma=1/3)$ throughout the period after inflation. In this
case, most of the PBHs will form at the epoch $t_1$ and have the mass
$M_1 =\eta (t_1 /t_{\rm Pl})\massp$ indicated by
Eq. (\ref{softmass}). This is because Eq. (\ref{PBHform}) implies
that, for a blue spectrum, $\beta_0 (M)$ decreases exponentially for
$M>M_1$, whereas PBHs with $M<M_1$ (which form before $t_1$) are
diluted by inflation. Thus we can regard the PBHs as effectively
having a $\delta$-function mass spectrum, as in the Barrow et
al. bubble collision scenario \c{BCL1992}, with a mass $M_1$. For
$\Gamma^{-1} \gg t_1$, the scalar field undergoes coherent
oscillations and the equation of state may be soft $(\gamma =0)$ from
$t_1$ until reheating is completed at the time $t_2 \equiv
\Gamma^{-1}$, when the scalar field decays rapidly into relativistic
particles \c{T1983}. In this case, the PBHs would have an extended
mass spectrum, going from $M_1$ to the mass indicated by
Eq. (\ref{maximummass}).

The quantum evaporation of PBHs by thermal emission leads to numerous
upper limits on the fraction of the Universe going into PBHs at a
given time. In the case of a radiation equation of state, upper limits
on $\beta_0 (M)$ in the range $10^{10}{\rm g}\le M\le 10^{17}$g have
been summarized in Refs. \c{CL1993,C1975,NPSZ1979} and are shown in
Figure (1a). The constraints on the probability of PBH formation are
modified if there is a dust phase after inflation (between $t_1$ and
$t_2$) because the ratio of PBH density to radiation density no longer
increases in this period. The constraints on the fraction $\beta (M)$
of the Universe going into PBHs during the dust phase are related to
$\beta_0 (M)$ via the equation \c{polnarev}
\be
\l{link}
\beta (M) =\beta_0 (M) \eta^{1/2} \left( \frac{t_2}{t_{\rm Pl}}
\right)^{1/2} \left( \frac{M}{\massp} \right)^{-1/2} .
\ee
This allows the limits on PBH formation during the dust phase to be
calculated from the constraints which apply if there is no dust
phase. The limits on $\beta (M)$ are indicated in Figures (2a), (3a)
and (4a), corresponding to three different choices of
$t_2$. Comparison of these observational limits with the predicted
values of $\beta_0 (M)$ and $\beta (M)$, given by Eqs. (\ref{PBHform})
and (\ref{betaM}), then leads to constraints on the form of $\delta
(M)$, as shown in Figures (1b), (2b) (3b) and (4b). The constraints on
$\beta_0(M)$, $\beta(M)$ and $\delta(M)$ for $M>10^{17}$g are
associated with non-evaporating PBHs and correspond to the requirement
that they have less than the critical density. We also show the limit
on $\delta(M)$ associated with the lack of spectral distortions in the
CMB implied by the \COBE  results \c{BC1991,D1991}; this limit is
now stronger than indicated in Ref. \c{CL1993}, as pointed out
independently by Hu et al. \c{FIRAS}. The constraints for
$M<10^{10}$g are associated with Planck mass relics and are the focus
of the rest of the paper.

\subsection{Formation of Relics}

The usual assumption is that evaporation proceeds until the PBH
vanishes completely \c{H1974}  but there are various arguments against
this. For example, the Uncertainty Principle implies that the
uncertainty in the mass $M$ of a Schwarzschild black hole of radius
$r_S=2GM/c^2$ must satisfy $c\Delta M \Delta r_S >\hbar /2$ and a
black hole evaporating below the Planck mass $\massp =(\hbar
c/G)^{1/2}$ would violate this \c{M1984}. Also the expected energy of
the emitted particle exceeds the rest mass of the black hole for
$M<\massp$. For these reasons Zel'dovich proposed that black holes
smaller than the Planck mass should be associated with stable
elementary particles \c{Z1984}. Another argument \c{BCL1992} is that
quadratic curvature corrections to the gravitational Lagrangian would
change the Hawking temperature to the form $T_{\rm BH}=k_1 M^{-1} -k_n
M^{-n}$, where $k_1$ and $k_n$ are constants, and this becomes zero
(implying that evaporation stops) once the mass gets down to
\be
\l{minmass}
M_{\rm rel} = \left({k_n \over k_1}\right)^{1/(n-1)}\massp .
\ee

Although we do not know what form quantum gravity corrections should
take as one approaches the Planck mass, it would be surprising if
there were none at all. The formation of relics is also related to the
paradox of information loss \c{H1974,infoprob}. The evaporation of a
black hole involves an initially pure quantum state evolving into a
mixed one and the basic principle of unitarity is thereby violated.
To avoid such a conclusion, one must either suppose that the
information is contained in the evaporated particles \c{Page} or that
the evaporation terminates when the black hole reaches the Planck
mass, the information being stored in a stable or very long-lived
relic
\c{stable}.

The above arguments are too vague to be very convincing, but more
specific arguments for stable relics have been given and we now review
some of these.

Bowick et al.  suggest that stable relics may form because black holes
can carry axionic charge \c{BGHHS}. The axionic field does not
gravitate and this means that a black hole may have arbitrarily large
charge without forming a naked singularity. On the other hand, it does
have a non-zero potential and this would be relevant for string
interactions (cf. the Aharonov-Bohm effect). This suggests that axion
charge will become important once a black hole has evaporated down to
of order the Planck mass since string effects will then
apply. Causality and energy conservation presumably limit the amount
of axion charge that a black hole can radiate within the age of the
universe, so Bowick et al. infer that it cannot evaporate completely.

Coleman et al. also argue for a minimum black hole mass on the basis
that a black hole can have quantum hair which affects its temperature
even though it has no classical effects \c{CPW1991}. They focus on
hair associated with a $Z_N$ gauge symmetry. Although a hole with
$Z_N$ electric charge has no {\em classical} hair, quantum effects
(associated with virtual strings that wrap around the event horizon)
generate a non-vanishing electric field which decays exponentially
with distance from the hole. In fact, this has negligible effect on
the temperature and much more important are the effects of {\em
magnetic} charge. This possibility arises if one considers $SU(N)/Z_N$
gauge theories which permit monopoles since a black hole can then have
classical magnetic hair. In this case, the Hawking process stops when
the temperature reaches zero and this occurs at a mass
\be
M_{\rm rel} = [n(N-n)/2N]^{1/2}e^{-1}\massp \approx 10^3\massp ,
\ee
where $n=1,2,\ldots, N-1$ is the $Z_N$ charge and $e$ is the coupling
constant. These relics can still decay to elementary monopoles if the
monopoles are light enough. Indeed the elementary monopoles may
themselves be $n=1$ black holes, in which case the extreme magnetically charged
black holes for other values of $n$ are kinematically forbidden to
decay to lighter objects of the same charge \c{LNW1992}. Thus the relics are
stable.

Gibbons and Maeda consider scale-invariant theories in which Maxwell
fields and antisymmetric tensor fields are coupled to gravity with a
dilaton \c{GM1988}. The solutions in four dimensions are of the
Reissner-Nordstr\"{o}m type, in which the black hole temperature goes
to zero when the mass equals the charge. For a general dimensionality
$D$, they find that the thermodynamic behaviour of an electrically
and/or magnetically charged black hole depends on $g$, the coupling of
the dilaton to the Maxwell fields. For $g>\sqrt{D-3}$ the black hole
evolves to a naked singularity with infinite temperature as in the
standard picture. For $g<\sqrt{D-3}$ a singular zero-temperature
endpoint is reached with a mass dependent on the scalar charge and of
order $g^{-1} \massp$. For $g=\sqrt{D-3}$ the hole evolves to a finite
temperature final state. It is interesting that string theory predicts
$g=1$ which corresponds to $D=4$ in this case.

Torii and Maeda consider a theory that couples a scalar dilaton field
to a Yang-Mills field \c{TM1993}. It arises as the 4-dimensional
effective theory corresponding to higher dimensional unified
theories. For coloured black holes with zero dilaton coupling they
find the effective Yang-Mills charge increases as the mass
decreases. The temperature also increases at low mass and becomes
infinite at a mass of $0.83M_{\rm Pl}/g_c$, where $g_c$ is the
coupling constant for the Yang-Mills field. At this point the event
horizon disappears leading to a stable particle-like solution. Similar
behaviour occurs for a sufficiently small dilaton coupling. They also
consider Skyrme black holes and find a similar effect. The evaporation
shrinks the event horizon until it disappears, leaving a non-Abelian
particle-like solution. The Skyrmion mass is determined by the
coupling constants and is of order $f_s g_c$, where $f_s$
relates to the mass of the Yang-Mills field \c{TTM1994}.

Callan et al. examine field equations which arise from low energy
string theory \c{CMP1988}. The action includes terms quadratic in the
curvature tensor, in addition to the usual Einstein action, so
Eq. (\ref{minmass}) is applicable. Solving the equations of motion by
perturbation expansion, they find that close to the black hole the
dilaton field decreases and the string interactions become weaker. The
stringy black hole temperature is unchanged for $D=4$, but it is lower
than the usual Hawking temperature for $D \ge 5$ by a factor
proportional to the inverse string tension. As such black holes lose
mass, their temperature reaches a maximum before falling to zero at a
finite mass.

Myers and Simon study black hole thermodynamics in second-order
theories in higher dimensions \c{MS1988}. Here the Lagrangian is the
sum of $k$ combinations of the Riemann invariants that give
second-order field equations. When $D=2k+1=5$, they find that the
temperature can vanish at finite mass. Although the horizon also
vanishes at this point, it requires an infinite amount of time to
evaporate to this limit and a naked singularity does not form. Whitt
also considers second-order gravity theories and obtains similar
results \c{W1988}.

\section{Observational Constraints from PBH Relics}

\setcounter{equation}{0}

In this section we will derive the constraints which can be placed on
the fraction of the Universe going into PBHs of mass $M$ in order to
avoid their relics having more than the critical density. We will
assume that the relics have a mass $\kappa \massp$, where the parameter
$\kappa$ is in the range $1-10^3$ for the scenarios discussed
above. We note at the outset that one can place no constraints on PBHs
which form before the end of inflation at $t_1$ since they will have
been diluted away. Thus there are no limits for PBHs with mass below
$M<M_1 =\eta (t_1/t_{\rm Pl})\massp $. Although we do not know the
value of $t_1$ {\em a priori}, we can place an upper limit $T_{\rm
max}$ on the reheat temperature and thus $T_1$ from the observed CMB
quadrupole anisotropy. Thus there are no constraints in the mass range
\be
\l{Mlessthan}
M<M_{\rm min} =\eta \massp (T_{\rm max}/T_{\rm Pl} )^{-2} .
\ee
If one assumes that the quadrupole anisotropy is due entirely to
tensor (gravitational wave) fluctuations, and that the spectral index
$n$ does not deviate significantly from unity, the expansion rate of
the Universe during inflation cannot exceed $H=2.9\times
10^{-5}\massp$ \c{Liddle,GGV1993}. This leads to a maximum reheat
temperature of $1.6\times 10^{16}$ GeV and a minimum mass $M_{\rm min}
\approx 2$g. A more careful determination of $M_{\rm min}$, allowing
for the contribution of the scalar fluctuations to the quadrupole
anisotropy, is given in Section 4 [c.f. Eq. (\ref{mmin})].

\subsection{Constraints for a Hard Equation of State}

Let us first consider the $\Gamma^{-1} \ll t_1$ situation, in which
the early Universe is radiation-dominated immediately after inflation.
The form of the relic constraint depends crucially on whether the PBHs
dominate the density before they evaporate at \c{H1974}
\be
\l{tevap}
t_{\rm evap}(M)=(M/\massp )^3 t_{\rm Pl} .
\ee
Since the ratio of the PBH density to radiation density increases as
$t^{1/2}$, the condition for the radiation to dominate at evaporation
is
\be
\l{evaporation}
\frac{\beta_0}{1-\beta_0} <\left( \frac{t}{t_{\rm evap}} \right)^{1/2}
=\eta^{-1/2} \left( \frac{M}{\massp} \right)^{-1} ,
\ee
where $t$ is the time at which PBHs of mass $M$ form and is  given by
Eq. (\ref{softmass}). This assumes that there are no PBHs in a
different mass range; we are essentially taking the PBHs to have a
$\delta$-function mass spectrum. In this case, the ratio of the relic
density to the critical density at the present epoch is
\be
\l{omegarel}
\Omega_{\rm rel} = \left( \frac{\beta_0}{1-\beta_0} \right) \left(
\frac{\kappa \massp}{M} \right) \left( \frac{t_{\rm eq}}{t} \right)^{1/2} ,
\ee
where $t_{\rm eq}=t_0 \Omega_{\rm rad}^{3/2}$ is the time at which the
matter and radiation densities are equal and $t_0=6.5h^{-1}$Gyr is the
age of the Universe for $\Omega_0=1$. Here $\Omega_{\rm rad}=2\times
10^{-5}h^{-2}$ is the current radiation density parameter and $h$ is
the Hubble parameter in units of $100$ km ${\rm s}^{-1}$ ${\rm
Mpc}^{-1}$. Using Eq. (\ref{softmass}), the constraint $\Omega_{\rm
rel}<1$ becomes
\be
\l{betalimit}
\frac{\beta_0}{1-\beta_0} < \eta^{-1/2}\kappa^{-1} \left(
\frac{t_0}{t_{\rm Pl}} \right)^{-1/2} \Omega_{\rm rad}^{-3/4}
\left( \frac{M}{\massp} \right)^{3/2} = 10^{-27}\eta^{-1/2}
\kappa^{-1} h^{2} \left( \frac{M}{\massp} \right)^{3/2} .
\ee
The PBHs dominate the density at $t_{\rm evap}$ if condition
(\ref{evaporation}) is not satisfied. In this case, most of the
background photons derive from the PBHs, so Eq. (\ref{omegarel}) is
replaced by
\be
\l{10}
\Omega_{\rm rel} = \kappa \left( \frac{\massp}{M} \right) \left(
\frac{t_{\rm eq}}{t_{\rm evap}} \right)^{1/2} = \kappa \left(
\frac{M}{\massp} \right)^{-5/2} \left( \frac{t_{\rm eq}}{t_{\rm Pl}}
\right)^{1/2}
\ee
and the constraint $\Omega_{\rm rel}<1$ becomes
\be
\l{Mbiggerthan}
M>\kappa^{2/5} \left( \frac{t_0}{t_{\rm Pl}} \right)^{1/5}
\Omega_{\rm rad}^{3/10}\massp =10^{11}\kappa^{2/5} h^{-4/5}\massp .
\ee
Note that this limiting mass also corresponds to the intersect of
conditions (\ref{evaporation}) and
(\ref{betalimit}). Eqs. (\ref{Mlessthan}), (\ref{betalimit}) and
(\ref{Mbiggerthan}) define the shaded lines on the left-hand-side of
Figure (1a). These equations were implicitly derived by Barrow et
al. \c{BCL1992}, although they expressed them in terms of $T$ rather
than $M$. {There is also a mistake in their Eq. (5.20) and this leads
to an error in their Eq. (5.31) which is equivalent to our
Eq. (\ref{Mbiggerthan}).

The relics have the critical density only on the boundary specified by
Eqs. (\ref{betalimit}) and (\ref{Mbiggerthan}).  It is interesting
that any value of $\beta$ above $10^{-12}$ will suffice to provide the
critical density if $M$ is fine-tuned to have the value given by Eq.
(\ref{Mbiggerthan}). Recall that most of the PBHs {\em actually} form
at the end of inflation, i.e. with the mass given by Eq.
(\ref{Mlessthan}), so this corresponds to fine-tuning the reheat
temperature to $T=10^{-5} \eta^{1/2} \kappa^{-1/5}T_{\rm Pl} \approx
10^{14}$GeV. Although one might regard this as unlikely, all the
present-day radiation originates from PBH evaporations in this
situation, so in this case one might regard $\Omega_{\rm rad}$ as a
free parameter determined by the reheat temperature.

Two other constraints are shown on the left of Figure (1a). The
stronger limit (shown dotted) comes from the fact that the observed
baryon asymmetry of the Universe could not be generated below the
critical temperature $T_{\rm min} ={\cal{O}}(10^3)$GeV associated with the
electroweak phase transition. PBHs that form with a mass in excess of
\be
\l{xs}
\massp (T_{\rm min} /T_{\rm Pl} )^{-2/3}=10^6{\rm g}
\ee
evaporate after this, so, if such holes come to dominate the Universe
before evaporation, the evaporated radiation is not sufficiently hot
to allow baryogenesis to proceed. This implies that condition
(\ref{evaporation}) must apply for $M\ge 10^6$g and this corresponds
to the dotted line in Figure (1a). Coincidentally, the masses given by
Eqs. (\ref{Mbiggerthan}) and (\ref{xs}) are nearly the same, so this
would exclude the critical density condition associated with
Eq. (\ref{Mbiggerthan}). However, this limit is not completely secure
since there may be other mechanisms (including black hole evaporations
themselves  \c{BAROPBH}) for generating baryon-asymmetry. A weaker but
more reliable constraint corresponds to the requirement that
evaporating PBHs do not generate a photon-to-baryon ratio exceeding
the current value $S_0 = 10^9$ \c{ZS1976}. One can show that this just
corresponds to the condition
\be
\frac{\beta_0}{1-\beta_0} < 10^9 \left( \frac{M}{\massp} \right)^{-1}
\qquad (M<10^{11}{\rm g}) ,
\ee
where the upper limit on $M$ arises because the PBHs must evaporate
early enough for the photons to be thermalized. This is weaker than
condition (\ref{evaporation}) by the factor $S_0$ and is labelled
`entropy' in Figure (1a).

In order to convert these constraints on $\beta_0 (M)$ into
constraints on the horizon-scale density fluctuations $\delta (M)$,
we use Eq. (\ref{PBHform}) to obtain
\be
\l{limitnorad}
\delta (M) < 0.16 \left[ \log10 \delta (M) - \log10 \beta_0 (M)
\right]^{-1/2} ,
\ee
where the value of $\beta_0$ at a given $M$ is determined by the
boundary of the shaded regions in Figure (1a). The relic constraint
then becomes
\bea
\l{17}
\delta (M) < 0.16 \left[ 27 -\log10 (\kappa\eta^{1/2})+1.5 \log10
\left( \frac{M}{\massp} \right) +\log10 \delta (M) \right]^{-1/2}
\nonumber \\ = 0.13 \left[ 17 -\log10 \left( \frac{M}{\massp}
\right) \right]^{-1/2} ~~~~~~~
\hbox{for $M<10^{11} \kappa^{2/5} h^{-4/5}\massp$,}
\eea
where we assume $\kappa=\eta =1$
and take $\log10 \delta (M) \approx -1.5$ in the second
expression. The relic limit on $\delta (M)$ is shown on the left of
Figure (1b).

\vspace{.35cm}
\centerline{\bf Figures 1a \& 1b}
\vspace{.35cm}

\subsection{Constraints for a Soft Equation of State}

We now consider the case in which the Universe has a dust era between
$t_1$ and $t_2$. We first assume that this ends before the PBHs
evaporate, so that
\be
\l{assump}
\frac{M}{\massp} > \left( \frac{t_2}{t_{Pl}} \right)^{1/3}
\ee
from Eq. (\ref{tevap}). The condition that the radiation dominates the
density when the PBHs evaporate now becomes
\be
\l{raddom}
\frac{\beta}{1-\beta} <\left( \frac{t_2}{t_{\rm evap}} \right)^{1/2}
= \left( \frac{t_2}{t_{\rm Pl}} \right)^{1/2} \left( \frac{M}{\massp}
\right)^{-3/2}
\ee
with no dependence on the time at which the PBHs form and the
condition $\Omega_{\rm rel} <1$ corresponds to the upper limit
\be
\l{anotherlim}
\frac{\beta}{1-\beta} < 10^{-27} \kappa^{-1}h^{2}\left( \frac{M}{\massp}
\right) \left( \frac{t_2}{t_{\rm Pl}} \right)^{1/2} .
\ee
Eqs. (\ref{raddom}) and (\ref{anotherlim}) just come from
Eqs. (\ref{evaporation}) and (\ref{betalimit}) together with
Eq. (\ref{link}). If the PBHs start to dominate the density before
$t_{\rm evap}$, then Eqs. (\ref{10}) and (\ref{Mbiggerthan}) still
pertain.

Next we assume that the dust era ends after $t_{\rm evap}$, so that
Eq. (\ref{assump}) is violated and the density of the
post-inflationary Universe is never dominated by radiation before
$t_2$. The current relic-to-radiation ratio therefore becomes
\be
\Omega_{\rm rel} = \beta \kappa \left( \frac{\massp}{M} \right)
\left( \frac{t_{\rm eq}}{t_2} \right)^{1/2}  \left[ (1-\beta) +\beta
\left( \frac{t_{\rm evap}}{t_2} \right)^{2/3} \right]^{-1} ,
\ee
where the two terms in square brackets give the contributions to the
radiation density from the reheating and the PBH evaporations. Since
the latter can be neglected, the condition $\Omega_{\rm rel}<1$ is
again given by Eq. (\ref{anotherlim}).

In order to determine the general constraints on $\beta (M)$ and
$\delta (M)$ one needs to specify the epoch $t_2$ at which the dust
phase ends. Eq. (\ref{maximummass}) then determines the mass range of
PBHs forming during the dust era. For $M_{\rm min}\le M\le M_{\rm
max}$ the constraint on $\beta (M)$ is given by Eq. (\ref{link}),
where the value of $\beta_0(M)$ is determined by the shaded boundary
in Figure (1a). Eq. (\ref{betaM}) then implies the upper limit
\be
\l{limitrad}
\delta (M)< 1.8 [\beta_0 (M)]^{2/13} \left( \frac{t_2}{t_{\rm Pl}}
\right)^{1/13} \left( \frac{M}{\massp} \right)^{-1/13} \eta^{1/13}
\ee
and, in particular, the relic limit given by Eq. (\ref{betalimit}) implies
\be
\delta (M) <1.3\times 10^{-4} \kappa^{-2/13} h^{4/13}\left( \frac{M}{
\massp}\right)^{2/13}\left( \frac{t_2}{t_{\rm Pl}} \right)^{1/13} .
\ee
The relic limits on $\beta (M)$ and $\delta (M)$, together with the
previously known constraints, are shown in Figures (2), (3) and (4).
Figure (2) applies if the dust era ends with the formation of
$10^{6}$g PBHs ($t_2 =10^{-31}$s), corresponding to the mass given by
Eq. (\ref{Mbiggerthan}). Figure (3) applies if it extends until
$10^{-22}$s, which is just after the $10^{10}$g PBHs form. Figure (4)
applies if it extends until $10^{-18}$s, which is just after the $10^{15}$g
PBHs form. [Eq. (\ref{maximummass}) implies that the condition for
this is $t_2>10^{-23} \delta(10^{15}{\rm g})^{-3/2}$s; this exceeds
the usual time of $10^{-23}$s because the PBHs are now smaller than
the particle horizon at formation.] Eqs.~(\ref{Mlessthan}),
(\ref{Mbiggerthan}) and (\ref{anotherlim}) define the shaded lines on
the left-hand-side of Figures (2a), (3a) and (4a). Qualitatively, the
effect of softening is to bring down the constraints relative to the
hard case. For low values of $t_2$, only the relic constraint is
brought down, as indicated in Figures (2). As $t_2$ increases, the
deuterium and gamma-ray limits are also brought down, as indicated in
Figures (3) and (4).

\vspace{.35cm}
\centerline{\bf Figures  2, 3 \& 4}
\vspace{.35cm}

\section{Results for a Constant Spectral Index}

\setcounter{equation}{0}

\subsection{Normalization to the Quadrupole}

So far the analysis has been completely general and has made no
assumptions about the form of the power spectrum of the density
fluctuations. In this section we shall consider the special case in
which the spectral index $n$ is constant and consider the
corresponding upper limit on its value. For a general inflaton
potential, $n$ will not be {\em exactly} constant but we show in the
Appendix that this  is always a good approximation in the hybrid
inflationary  scenario. The strongest upper limit on $n$ is derived by
normalizing at the \COBE  quadrupole scale and finding the steepest
straight line which avoids all the shaded areas in Figures (1b), (2b), (3b)
and (4b). In case (1), the strongest constraint is associated with the
relics from the PBHs of mass $M_{\rm min}$. This means that relics can
have the critical density but that PBHs cannot contribute appreciably
to $100$MeV cosmic rays. In case (2), the constraint on $n$ associated
with the relics is even stronger because only the relic dip is brought
down by the dust era. However, in case (3), the deuterium limit gives
the strongest constraint on $n$ because the $10^{10}$g dip has also
been pulled down. In case (4) the gamma-ray limit gives the strongest
constraint because the $10^{15}$g dip has been pulled down, so PBHs
can generate the cosmic rays but not a critical density of
relics. Note that it is not possible to produce {\em both} the cosmic
rays and a critical relic density, i.e. there is no situation in which
a single straight line passes through both the $M_{\rm min}$ and
$10^{15}$g points. Either the $10^{15}$g PBHs form during the dust
era, in which case the gamma-ray limit implies the relics are
unimportant, or they form after the dust era, in which case the relic
limit implies the gamma-ray limits  are unimportant.

In order to determine the constraint on $n$, we need to normalize the
density spectrum at the quadrupole scale. To do this, we must first
convert the observed temperature fluctuation into a corresponding
density fluctuation.  Inflation also produces primordial tensor
fluctuations (gravity waves) and these can be significant in some
cases. If no reionization occurs, the surface of last scattering is
located at a redshift $z_{\rm LS} \approx 1100$ for $\Omega_{0}=1$ and
the angle subtended by the horizon at that redshift is approximately
$\theta \approx z_{\rm LS}^{-1/2} \approx 2^o$.  Once gravity waves
reenter the horizon, they redshift as relativistic matter and soon
become negligible.  However, experiments probing angular scales larger
than $2^o$ measure superhorizon-sized perturbations and are therefore
sensitive to gravitational wave effects.

The CMB anisotropies can be expanded into spherical harmonics with
coefficients $a_{lm}({\bf x})$ which are stochastic random variables.
The observed multipoles detected from a single point in space are
defined in terms of these coefficients: $Q^2_l=\frac{1}{4\pi}
\sum^l_{m=-l}|a_{lm}|^2$ \c{SV1990}. A given inflationary model
predicts values for the averaged quantities $\Theta_l^2 =
\frac{1}{4\pi} (2l+1) \langle |a_{lm} ({\bf x}) |^2
\rangle$, where the average is taken over all observer points. If the
$a_{lm}({\bf x})$ are uncorrelated, the expected contributions from
the scalar $(\Theta^2_l[S])$ and tensor $(\Theta_l^2[T])$ fluctuations
add in quadrature, so $\Theta^2_l=\Theta^2_l[S]+\Theta^2_l[T]$.

We follow the notation of Ref. \c{Turner} and define the scalar and
tensor contributions to $\Theta_2^2$ as
\be
S\equiv \Theta_2^2 [S] = \frac{5\langle |a_{2m}[S]|^2 \rangle}{4\pi}, \qquad
T\equiv \Theta_2^2 [T] = \frac{5\langle |a_{2m}[T]|^2\rangle}{4\pi} ,
\ee
respectively.  The observable quantities $S$ and $T$ have been
calculated numerically in terms of the corresponding values of the
potential and its derivatives \c{Turner}. The relevant expressions for
the present work are
\be
\l{V}
\frac{V_C}{\massp^4} \approx 1.65 \left( 1+0.20 \frac{T}{S} \right) T
\ee
\be
\l{V'}
\frac{|V_C'|}{\massp^3} \approx  3.14 \frac{T^{3/2}}{S^{1/2}} ,
\ee
where a subscript $C$ refers to the quadrupole scale. Substituting
these expressions into Eq. (\ref{scalar}) yields the magnitude of the
density fluctuation at the quadruole:
\be
\l{densityquad}
\delta_C \approx 0.67 S^{1/2} \left( 1+0.3 \frac{T}{S} \right) .
\ee

\subsection{Constraints on the Spectral Index for no Dust Phase}

If the Universe has the critical density, the comoving rest mass
within a sphere of radius $\lambda /2$ is $M=1.5\times
10^{11}M_{\odot}h^2(\lambda /{\rm Mpc})^3$. The quadrupole corresponds
to a scale $\lambda_C \approx 6000  h^{-1}$ Mpc and has an
associated mass $M_C=  10^{57}h^{-1}$g.  For a power law
spectrum we require $\delta_C (M/M_C)^{(1-n)/6}$ be less than the
value of $\delta (M)$ indicated in Figures (1b)-(4b). Taking
logarithms and substituting Eq. (\ref{densityquad}) then yields an
upper limit on $n$ in terms of the $S$ and $T$:
\be
\l{constraint}
n-1 \le 6 \left[ {\rm log}_{10} \left(  \frac{  10^{57} h^{-1}
{\rm g}}{M} \right) \right]^{-1} \left[ 0.17 +\log10 \left(
\frac{\delta (M) }{\sqrt{S}} \right)    -  \log10 \left( 1 +0.3\frac{T}{S}
\right) \right] .
\ee
The last logarithmic term is a correction accounting for the
gravitational wave contribution to the observed quadrupole anisotropy.
Since $\log10 (\delta / S^{1/2}) \ge 4$, we conclude that this
correction term is negligible. Furthermore, although the \COBE
experiment measures the quantity $S+T$, it is easily seen that limit
(\ref{constraint}) is not altered significantly if it is assumed that
the entire \COBE signal is due to the scalars.  This shows that a
separate determination of the gravitational wave contribution is not
required in order to derive a limit on $n$.

A second feature of this limit is its relative insensitivity to cosmic
variance. Cosmic variance arises because the fluctuations predicted by
theory are stochastic in nature and have a Gaussian probability
distribution. A set of observations can only measure a finite number
of realizations of this distribution and this is never sufficient to
specify it completely. Thus there always exists an intrinsic
uncertainty in the analysis. Although the effect is most significant
at the quadrupole scale and can be problematic when normalizing
large-scale structure observations, it should not significantly affect
the PBH limit since this spans such a large range of scales.

It only remains to substitute the numerical value of $S^{1/2}$ into
Eq. (\ref{constraint}). A maximum likelihood estimation made by Seljak
and Bertschinger \c{SB1993} with the first-year \COBE data implies a central
value of
\be
\l{theta2}
S^{1/2} = 5.8 \times 10^{-6} e^{0.46(1-n)}
\ee
if the blackbody temperature of the CMB radiation is $T_{\rm rad}
=2.736$K. Smoot et al. \c{S1994} and Bennett et al. \c{Be1994} arrive
at similar expressions, although we emphasize that the precise form of
the relationship between $S$ and $n$ is not important for the PBH
constraints. Hence, substituting Eq. (\ref{theta2}) into Eq.
(\ref{constraint}) implies
\be
\l{mainconstraint}
n-1 \le \frac{6}{56-\log10 (M/{\rm g})} \left[ 5.4 +\log10 \delta (M)
\right] \qquad (M\ge M_{\rm min}) .
\ee

An expression for $M_{\rm min}$ may be derived in terms of quantities
that are in principle observable. During reheating the potential
energy of the scalar field driving inflation is converted into
relativistic particles with an energy density $\rho_{\rm rad} =\pi^2
\epsilon T_1^4/30$, where $\epsilon$ is the number of relativistic
degrees of freedom. Typically $\epsilon = {\cal{O}} (10^2)$ and
$\epsilon \approx 160$ in the simplest SU(5) GUT model. An observable
upper limit on the reheat temperature may be calculated by assuming
that the available energy is given by $V_C$. For an efficient
reheating process this gives
\be
\l{t_one}
T_1 \le \left( \frac{30}{\pi^2\epsilon} V_C \right)^{1/4}
 =  \left[ \frac{5}{\epsilon} \left( 1+0.2 \frac{T}{S} \right) T \right]^{1/4}
T_{\rm Pl} ,
\ee
where the second expression uses the lowest order terms in Eq.
(\ref{V}).  The smallest black holes formed after inflation must
therefore have a mass exceeding
\be
\l{mmin}
M_{\rm min} = 1.7 \times 10^{-6} \left( \frac{\epsilon}{T} \right)^{1/2} \left(
1-0.1 \frac{T}{S} \right) {\rm g} .
\ee
 In principle one requires a knowledge of the
tensor contribution to determine $M_{\rm min}$ but such information is
not currently available.

In the efficient reheating case, the strongest upper bound on the
spectral index $n$ depends on the value of $M_{\rm min}$ or
equivalently $t_1$. For $M_{\rm min} < 10^6$ g,
the strongest limit is associated with the relics and the relevant
mass is $M_{\rm min}$ itself. Substituting Eq. (\ref{17}) into Eq.
(\ref{mainconstraint}) then gives
\be
\l{relicconstraint}
n-1 \le \frac{6}{56 -\log10 (M_{\rm min}/{\rm g})} \left[ 4.5 -\frac{1}{2}
\log10 \left( 12 -\log10 \left( \frac{M_{\rm min}}{\rm g} \right)
\right) \right] .
\ee
Thus the limit on the spectral index is $n \le 1.4$ for $M_{\rm
min}\approx 1$g, corresponding to $T_1={\cal{O}}(10^{16})$GeV, and $n
\le 1.5$ for $M_{\rm min}\approx 10^6$g, corresponding to $T_1
={\cal{O}}(10^{14})$GeV. The limits on $n$ for other ranges of $M_{\rm
min}$ are shown by the upper shaded curve in Figure 5. Here we are
regarding $n$ as a continuous function of $T_{1}$ (or equivalently
$t_{1}$). The closure density constraint applies for $M_{\rm min} \ge
10^{17}{\rm g}$ ($t_1 /t_{\rm Pl} \ge 10^{22}$), the gamma-ray
constraint for $10^{13}{\rm g}
\le M_{\rm min} \le 10^{17}{\rm g}$ ($10^{ 18} \le t_1  /t_{\rm Pl} \le
10^{ 22}$) and the deuterium constraint for $10^{10}{\rm g} \le M_{\rm
min} \le 10^{13}{\rm g}$ ($10^{ 15} \le t_1 /t_{\rm Pl} \le 10^{
18}$). Since the helium and entropy constraints are associated with
higher values of $\delta_{\rm H}$ than the deuterium constraint, the
limit on $n$ does not improve until $M_{\rm min}$ gets down to $10^6
{\rm g}$ when the relic constraint (\ref{relicconstraint}) becomes
relevant. This is why the part of the upper curve in Figure (5)
between ``relics'' and ``deuterium'' is flat, with a discontinuity at
$10^6$g. Note that the CMB distortion gives a better limit on $n$
than the closure density limit for $M_{\rm min} > 10^{24} g$ (or
$t_1 > 10^{-14}$s) and this is why the limit flattens off at $n=1.76$.
Hu et al \c{FIRAS} have calculated this limit somewhat more carefully
and find that it flattens off at n=1.54, as indicated by the dotted
line in Figure 5.

\subsection{Constraints for an  Early Dust Phase}

If there is a dust phase after inflation, the constraints on the index
$n$ depend on both the value of $t_1$ and $t_2$, with the latter
itself being determined by the decay width $\G$. We first assume that
$t_1$ is fixed at the value associated with Eq. (\ref{t_one}) and
allow $\G$ to vary. An upper limit on $\G$ follows from the fact that
the reheat temperature at the end of the dust phase is $T_{\rm RH}
\approx (\G t_{\rm Pl})^{1/2} \massp$. Therefore if efficient
conversion of the vacuum energy to relativistic particles occurs, the
assumption that the \COBE quadrupole signal is due entirely to
gravitational waves implies $\G t_{\rm Pl} \le 10^{-6}$ as shown by
the shaded line on the left of Figure 5. The more general limit is $\G
t_{\rm Pl} < (T_1 /T_{\rm Pl})^2$ where $T_1$ is given by
Eq. (\ref{t_one}). A lower limit on $T_{\rm RH}$ follows from the
requirement that baryogenesis must proceed after reheating. If the
lowest temperature for which the observed baryon asymmetry may be
generated is the electroweak scale, ${\cal{O}}(10^3)$GeV, we require
$\G t_{\rm Pl} \ge 10^{-30}$. As discussed in Section 3.1, there is
some uncertainty in the argument, so we show this limit by a broken
limit in Figure 5. For intermediate values of $\G$, one can merely
exclude certain areas in the $(\G,n)$ plane as we now demonstrate.

If we normalize on the \COBE quadrupole scale, $M_C$, the rms
amplitude on a smaller scale $M$ is $\delta (M) = \delta_C
(M/M_C)^{(1-n)/6}$ where $\delta_C \approx 3.8\times 10^{-6}$ from
Eqs. (\ref{densityquad}) and (\ref{theta2}). Substituting this
expression into Eq.  (\ref{maximummass}) implies that the maximum mass
of a PBH formed during the dust phase is
\be
\l{Mmax}
M_{\rm max} \approx \left(\frac{\delta_C^{3/2}}{\G t_{\rm Pl}}
\right)^{4/(n+3)} \left( \frac{M_C}{\massp}\right)^{(n-1)/(n+3)} \massp .
\ee
It follows that PBHs with mass $M$ are formed during the dust phase
only if
\be \l{Mlim}
\log10 \G t_{\rm Pl} \le -23.6 +15.5n
-\left( \frac{3+n}{4} \right)\log10 \left( \frac{M}{\massp}  \right) .
\ee
Another upper limit on $\Gamma t_{\rm Pl}$ follows from
Eq. (\ref{limitrad}) with $t_2$ identified with $\G^{-1}$:
\be
\l{Gamma}
\log10 \G t_{\rm Pl} \le 208 -134n +\frac{1}{6}
(13 n -19) \log10 \left( \frac{M}{\massp} \right) + 2\log10 \beta_0
(M)
+ \log10 \eta ,
\ee
where the value of $\beta_0(M)$ is indicated in Figure (1a). For a
given value of $\G$, Eqs. (\ref{Mlim}) and (\ref{Gamma}) can also be
interpreted as giving lower and upper limits on $n$.

The strongest limit on $n$ or $\G$ is associated with the relic
constraint for $M_{\rm max}<10^{10}$g, the deuterium constraint for
$10^{10}{\rm g}<M_{\rm max}<10^{15}{\rm g}$ and the gamma-ray
constraint for $M_{\rm max}>10^{15}$g. Eq. (\ref{Mmax}) implies that
the relic constraint applies for
\be
\l{relic}
\log10 \G t_{\rm Pl} > -35 + 12n
\ee
and Eqs. (\ref{betalimit}) and (\ref{Gamma}) with $M=M_{\rm min}\approx
1{\rm g}$ then imply
\be
\l{relrange}
\log10 \G t_{\rm Pl} < 153 - 125n.
\ee
Similarly the deuterium constraint, $\beta_0 (10^{10}{\rm g}) <
10^{-21}$, applies for
\be
\l{deut}
-35+12n>\log10 (\G t_{\rm Pl}) > -38.6 + 10.5n
\ee
and Eq. (\ref{Gamma}) with $M=10^{10}$g gives
\be
\l{deutlim}
\log10 (\G t_{\rm Pl}) < 118 -101n .
\ee
The gamma-ray constraint, $\beta_0 (10^{15}{\rm g})<10^{-26}$, applies
for
\be
\l{csraylim}
\log10 (\G t_{\rm Pl}) < -38.6 + 10.5n
\ee
and Eq. (\ref{Gamma}) with $M=10^{15}$g gives
\be
\l{cosmicray}
\log10 (\G t_{\rm Pl}) < 93 - 91n .
\ee
 These constraints on $(\G,n)$ are shown by the lower shaded line in
Figure (5).  The relic constraint applies for $10^{-8} \ge \G t_{\rm
Pl} \ge 10^{-17}$, the deuterium constraint for $10^{-17} \ge \G
t_{\rm Pl} \ge 10^{-23}$ and the gamma-ray limit for $10^{-23}\ge \G
t_{\rm Pl}
\ge 10^{-30}$.
We therefore arrive at an upper limit of $n=1.4$ if PBH form relics
and there is an extended dust phase immediately before the standard
radiation-epoch.  This limit is independent of the form of the dark
matter, the bias parameter, cosmic variance and any gravitational wave
contribution to the \COBE signal.  It is clear from Figure (5) that
the range $1.3<n<1.4$ is astrophysically very interesting. For $T_{\rm
RH} <10^6$GeV the origin of the observed gamma-ray and cosmic ray
spectra can in principle be explained in terms of evaporating
$10^{15}$g PBHs formed during the early dust phase. This corresponds
to the limit (\ref{cosmicray}) becoming an equality. For $T_{\rm RH}
\ge 10^{9.5}$GeV, however,  the Planck mass relics of 1g PBHs may be a natural
candidate for the cold dark matter in the universe. Note that if $t_1$
is allowed to increase  associated with Eq (\ref{t_one}), then the
limits on ($n,\G$) are somewhat weaker than indicated by the lower
curve. The lower curve therefore gives the most stringent limit.

\vspace{.35cm}
\centerline{\bf Figure 5}
\vspace{.35cm}

\section{Conclusions and Discussion}

\setcounter{equation}{0}

In this paper we have derived upper limits on the fraction of the
Universe going into PBHs at various mass scales and thereby inferred
upper limits on the rms amplitude of the initial density
perturbations. In the inflationary scenario the spectrum of density
fluctuations arising from quantum fluctuations in the vacuum is
uniquely determined by the functional form of the self-interaction
potential of the inflaton field, so it is possible to constrain the
inflaton potential with the PBH constraints.

One of our main purposes has been to explore the consequences of the
suggestion that evaporating PBHs may leave behind stable relics with
masses of order the Planck mass. The requirement that such relics have
less than the critical energy density leads to constraints which are
stronger than those previously derived for masses smaller than
$10^6$g. We have also investigated to what extent the PBH constraints
are altered if there was an early dust phase in the history of the
Universe. Generically inflation ends via a phase transition which can
be either first- or second-order. In the latter case, the scalar field
rolls down to the global minimum of its potential and begins to
undergo coherent oscillations. If the decay of the field to
relativistic particles is rapid relative to the expansion rate of the
Universe, reheating occurs almost instantaneously. However these
oscillations can be prolonged if the coupling constants are
sufficiently small and, in this case, the Universe becomes dust like
for some period before the standard radiation epoch is recovered. This
is very important for the PBH constraints because this is precisely
the period in which the PBHs are expected to form. The constraints now
depend crucially on the duration of this dust phase but the upper
limits on the rms amplitude are always strengthened.

In order to constrain the spectral index $n$ of the density
fluctuations, we have normalized the rms amplitude on the \COBE
quadrupole scale and applied the above results to the special case in
which the spectral index is constant over all the scales of interest.
This example has recently been investigated within the context of the
hybrid inflationary scenario. The limit on $n$ depends crucially on
the reheat temperature $T_{\rm RH}$. If there is no early dust phase,
the relic constraint provides the strongest upper limit for $T_{\rm
RH} > 10^{14}$GeV and this corresponds to PBHs with masses smaller
than $10^6$g. The limit is $n\approx 1.42$ for $T_{\rm RH} \approx
10^{16}$GeV and $n\approx 1.49$ for $T_{\rm RH} \approx 10^{14}$GeV.
The constraint is tightened if there was an early dust phase and it
was found that the range $1.3<n<1.4$ has a number of potentially
interesting astrophysical consequences. In particular relics can
provide the critical density for $T_{\rm RH} > 10^{9.5}$GeV and
$10^{15}$g PBHs can contribute appreciably to the observed gamma-ray
and cosmic-ray spectra at $100$MeV for $T_{\rm RH} > 10^6$GeV. It is
interesting that the current best-fits to the second-year \COBE data
appear to favour this range for the spectral index, although it must
be emphasized that these fits are also consistent with a flat $(n=1)$
spectrum. It is also intriguing that the BATSE detector on the Compton
Gamma Ray Observatory (CGRO) \c{CGRO} has observed gamma-ray bursts
that appear to exhibit the characteristics associated with PBH
explosions at the present epoch \c{CH1994}. It remains to be seen,
however, whether this range of values for the spectral index can be
reconciled with current data from large-scale structure.

A number of assumptions have been made in this work. For simplicity we
have assumed that the transition from the coherent oscillation phase
to the radiation phase occurs instantaneously, although this is not
necessarily the case \c{T1983}. Moreover, during the oscillation phase,
the scalar field only behaves as a dust fluid on average over one
cycle. The formation of PBHs in such a case may be different to that
assumed in Ref. \c{polnarev} and presumably this would
affect the observational limits. Moreover, in deriving the constraints
on the spectral index, we assumed that $n$ is constant over all
relevant scales.

In conclusion, therefore, the above constraints are useful for a
number of reasons. Firstly they extend to scales many orders of
magnitude smaller than those accessible to large-scale structure
observations. This implies that errors in the \COBE  detection due, for
example, to cosmic variance are not important when one normalizes the
spectrum on the  scales associated with large angle CMB
experiments. Likewise, any contribution to the large-scale CMB
anisotropy from a primordial spectrum of gravitational waves is
generally also negligible. Moreover, both the \COBE  point and the PBH
limits are independent of the form of the dark matter in the Universe
and the constraints on the spectral index are therefore independent of
any transfer functions and biassing parameters. Consequently, although
they are somewhat weaker than those derived from current large-scale
structure observations alone, they have the advantage that they are
relatively insensitive to any specific choice of cosmological model.

\vspace{1.5cm}

{\bf Acknowledgments} JHG and JEL are supported by the Science and
Engineering Research Council (SERC), UK. JEL is supported at Fermilab
by the DOE and NASA under Grant No. NAGW-2381. We would like to thank
R. C. Caldwell, A. R. Liddle, K. Maeda, A. Polnarev and D. Scott for fruitful
discussions regarding the contents of this work.
\\

\newpage

\section*{Appendix}

\normalsize
\setcounter{equation}{0}

\def\theequation{A.\arabic{equation}}

In the case of hybrid inflation, the constancy of the spectral index
is a natural result \c{hybrid,Cetal}. In this scenario one of the
scalar fields is initially at zero and the potential takes the
asymptotic form
\be
\l{A1}
V(0,\phi) = V_0 + {1\over2}m^2\phi^2 ,
\ee
where $V_0={M^4 / 4\lambda^2}$ and $M$ is a free parameter. Such
models assume $\lambda m^2 \ll M^2 g^2$ and $3\lambda m^2 \massp^2 \ll
2\pi M^4$, where $\lambda$ and $g$ are coupling constants typically of
order unity \cite{hybrid}. The inflationary phase ends when $\phi$
reaches $\phi_e = M/g$ and triggers the rapid rolling of the second
field. The spectral index and number of e-folds are given by
\cite{LL1992}
\bea
\l{n}
n = 1 + {\massp^2 \over 4\pi}{V''\over V} - {3\massp^2 \over 8\pi}
\left({V'\over V}\right)^2 \\
N = {8\pi \over \massp^2} \int_{\phi_e}^{\phi} {V \over V'} d\phi ,
\eea
where the quantities on the right hand side of Eq. (\ref{n}) are
evaluated when the scale first crosses the Hubble radius during
inflation. If one requires $n>1$, the false vacuum term $V_0$ in the
potential must dominate the $m^2 \phi^2$ contribution \cite{CL1993}.
This is the case near the end of inflation, so the changes in $n$ and
$N$ are given by
\be
\l{dn1}
\Delta n \simeq -{3\massp^2 \over 4\pi} {m^4 \phi \over V_0^2} \Delta \phi
\ee
and
\be
\Delta N \simeq {8\pi \over \massp^2} {V_0 \over m^2 \phi} \Delta \phi ,
\ee
respectively. Eliminating $\Delta \phi$ and substituting for $V_0$
with $\phi = \phi_e$ gives
\be
\l{dn2}
\Delta n \simeq - \Delta N {8\pi\lambda^3 \over 3}\left({m \massp
\over M^2} \sqrt{3\lambda \over 2\pi} \right)^4 {\lambda m^2 \over M^2 g^2}
\ee
and for $\Delta N \simeq 60$ this implies that  $\Delta
n \ll 1$ over the range of interest. We conclude, therefore, that $n$
is effectively constant over the last 60 e-foldings of inflation. This may be
understood physically by considering the form of the effective potential
(\ref{A1}) as   $\phi$  rolls towards the minimum. As  $\phi$ decreases the
form of Eq. (\ref{A1}) appro
aches the secant potential that leads to an exactly constant  spectral index.

\newpage
\vspace{2.5in}

{\em Figure 1}: (a) The constraints on the fraction $\beta_0 (M)$ of
the early Universe going into PBHs with mass $M$ if the equation of
state is radiation-like ($\gamma =1/3$). The origin of the constraints
above $10^{10}$g is summarized in Ref. \c{CL1993}. The ``entropy''
constraint arises from the requirement that the PBH evaporations do
not generate more than the observed photon-to-baryon ratio. There is
potentially a stronger but less secure constraint in this mass range
derived from the assumption that the observed baryon asymmetry must be
generated above the electroweak scale and this is shown dotted. Below
$10^6$g the strongest limit is due to the relics left over from PBH
evaporations.  (b) The corresponding constraints on the rms amplitude
of the density fluctuations.  If PBHs do not leave behind stable
relics after evaporation, the strongest upper bound on the spectral
index is given by the  dashed line which joins the \COBE point and
the deuterium constraint at $10^{10}$g. This limit applies for reheat
temperatures above ${\cal{O}}(10^9)$GeV. If relics are formed, the
limit is strengthened at higher reheat temperatures as indicated.

\vspace{.35cm}

{\em Figure 2}: (a) The constraints on $\beta(M)$, the fraction of the
Universe going into PBHs of mass $M$ during a post-inflationary dust
phase. The dust phase is assumed to last until $t_2 \approx
10^{-31}$s, long enough for $10^6$g configurations to detach
themselves from the cosmological expansion before the radiation era
begins. (b) The corresponding constraints on $\delta (M)$, with the
lines having the same significance as in Figure (1b). Although the
limit on the probability of PBH formation is reduced during the dust
phase, the upper limit on the rms amplitude of the fluctuation at a
given scale is increased because the fraction of the Universe going
into PBHs is no longer exponentially damped.

\vspace{.35cm}

{\em Figures (3a) and (3b)}: The same as Figures (2a) and (2b) with
the duration of the dust phase extended until $t_2 \approx 10^{-22}$s to
allow $10^{10}$g PBHs to form.

\vspace{.35cm}

{\em Figures (4a) and (4b)}: (a) The same as for Figures (2a) and (2b)
with the duration of the dust phase extended until $t_2 \approx
10^{-18}$s to allow $10^{15}$g PBHs to form.

\vspace{.35cm}

{\em Figure 5}: Illustrating the constraints on the spectral index
arising from the overproduction of primordial black holes, the shaded
area being excluded. The lower line applies if there is a dust phase
immediately after inflation, in which case the ordinate is $\log10 \G
t_{\rm Pl}$, the upper line if there is no dust phase in which case it
is $\log10(t_{\rm Pl}/t_1)$. The constraints depend on the reheat
temperature $T_{\rm RH} \approx 10^{18} (\Gamma t_P )^{1/2}$GeV, where
$\G$ is the decay width of the scalar field that decays into relativistic
particles. The $n$-independent upper and lower limits on
the decay width arise from assuming the
\COBE detection is due entirely to gravitational waves (shaded line) and from
requiring that baryogenesis can only proceed above the electroweak
scale (dashed line). The dotted horizontal  line indicates the CMB distortion
limit of Hu et al \c{FIRAS}. For reheat temperatures above $T_{\rm RH}
\approx 10^{9.5}$GeV the most important constraint arises from the
requirement that any Planck mass relics left over from the final
stages of PBH evaporation should have less than the critical density
at the present epoch. For lower reheat temperatures, more massive PBHs
may form and the strongest constraints then arise from the
photodissociation of deuterium by evaporating $10^{10}$g PBHs, from
the observed gamma-ray background in the energy range $0.1-1$ GeV, or
from the distortions of the CMB. It is clear from this figure that the
region $1.3\le n\le 1.4$ has a number of interesting astrophysical
consequences if there was an extended dust phase immediately after the
inflationary expansion.

\end{document}